\documentclass[preprint,journal]{vgtc}

\usepackage{booktabs} 
\usepackage{CJKutf8}

\usepackage{graphicx}
\usepackage{microtype}                  
\PassOptionsToPackage{warn}{textcomp}  
\usepackage{textcomp}                  
\usepackage{times}                     
\usepackage{tabu}                      

\usepackage{balance}  
\usepackage{times}    
\usepackage{url}      
\usepackage{enumitem}
\usepackage{xspace}
\usepackage[table,usenames,dvipsnames]{xcolor}
\usepackage{booktabs}
\usepackage{multirow}
\usepackage[10pt]{moresize} 
\usepackage{changepage}
\usepackage{amssymb}
\usepackage{amsmath}
\usepackage{hyperref}
\usepackage{amsthm}
\usepackage{xcolor}
\usepackage[compact]{titlesec}
\usepackage{soul}
\usepackage[colorinlistoftodos,prependcaption,textsize=tiny]{todonotes}

\newcommand{\del}[1]{} 

\vgtcpapertype{application/design study}

\usepackage{cellspace}

\theoremstyle{definition}

\onlineid{0}

\vgtccategory{Research}

\title{FairSight: Visual Analytics for Fairness in Decision Making}
\author{Yongsu Ahn, Yu-Ru Lin}
\authorfooter{
\item
 Yongsu Ahn is with University of Pittsburgh. E-mail: yongsu.ahn@pitt.edu.
\item
 Yu-Ru Lin is with University of Pittsburgh. E-mail: yurulin@pitt.edu.
}

\newcommand{\name}{\textsl{FairSight}\xspace}
\newcommand{\fairdm}{\textsl{FairDM}\xspace}

\newcommand{\data}{\fontfamily{lmss}\selectfont\textcolor{RedViolet}{Data}\normalfont\xspace}
\newcommand{\model}{\fontfamily{lmss}\selectfont\textcolor{RedViolet}{Model}\normalfont\xspace}
\newcommand{\outcome}{\fontfamily{lmss}\selectfont\textcolor{RedViolet}{Outcome}\normalfont\xspace}

\newcommand{\dataabbr}{\fontfamily{lmss}\selectfont\textcolor{RedViolet}{D}\normalfont}
\newcommand{\modelabbr}{\fontfamily{lmss}\selectfont\textcolor{RedViolet}{M}\normalfont}
\newcommand{\outcomeabbr}{\fontfamily{lmss}\selectfont\textcolor{RedViolet}{O}\normalfont}

\newcommand{\inputspace}{\fontfamily{lmss}\selectfont\textcolor{RedViolet}{Input}\normalfont\xspace}
\newcommand{\mappingspace}{\fontfamily{lmss}\selectfont\textcolor{RedViolet}{Mapping}\normalfont\xspace}
\newcommand{\outputspace}{\fontfamily{lmss}\selectfont\textcolor{RedViolet}{Output}\normalfont\xspace}

\newcommand{\understand}{\fontfamily{lmss}\selectfont\textcolor{Blue}{Understand}\normalfont\xspace}
\newcommand{\measure}{\fontfamily{lmss}\selectfont\textcolor{Blue}{Measure}\normalfont\xspace}
\newcommand{\identify}{\fontfamily{lmss}\selectfont\textcolor{Blue}{Identify}\normalfont\xspace}
\newcommand{\mitigate}{\fontfamily{lmss}\selectfont\textcolor{Blue}{Mitigate}\normalfont\xspace}

\newcommand{\understandabbr}{\fontfamily{lmss}\selectfont\textcolor{Blue}{U}\normalfont}
\newcommand{\measureabbr}{\fontfamily{lmss}\selectfont\textcolor{Blue}{M}\normalfont}
\newcommand{\identifyabbr}{\fontfamily{lmss}\selectfont\textcolor{Blue}{I}\normalfont}
\newcommand{\mitigateabbr}{\fontfamily{lmss}\selectfont\textcolor{Blue}{MT}\normalfont}

\newcommand{\fair}{\fontfamily{lmss}\selectfont\textcolor{OliveGreen}{Fair}\normalfont\xspace}
\newcommand{\decision}{\fontfamily{lmss}\selectfont\textcolor{OliveGreen}{Decision}\normalfont\xspace}
\newcommand{\explain}{\fontfamily{lmss}\selectfont\textcolor{OliveGreen}{Explain}\normalfont\xspace}

\newcommand{\generator}{Generator\xspace}
\newcommand{\rankingview}{Ranking View\xspace}

\newcommand{\globalinspection}{Global Inspection View\xspace}
\newcommand{\localinspection}{Local Inspection View\xspace}
\newcommand{\featureinspection}{Feature Inspection View\xspace}
\newcommand{\rankinglist}{Ranking List View\xspace}

\renewcommand{\vec}[1]{\mathbf{#1}}
\newcommand{\mat}[1]{\mathbf{#1}}

\newcommand{\topk}{top-$k$\xspace}
\abstract{
Data-driven decision making related to individuals has become increasingly pervasive, but the issue concerning the potential discrimination has been raised by recent studies. In response, researchers have made efforts to propose and implement fairness measures and algorithms, but those efforts have not been translated to the real-world practice of data-driven decision making. As such, there is still an urgent need to create a viable tool to facilitate fair decision making. We propose \name, a visual analytic system to address this need; it is designed to achieve different notions of fairness in ranking decisions through identifying the required actions -- understanding, measuring, diagnosing and mitigating biases -- that together lead to fairer decision making. Through a case study and user study, we demonstrate that the proposed visual analytic and diagnostic modules in the system are effective in understanding the fairness-aware decision pipeline and obtaining more fair outcomes.
}


\teaser{
  \centering
  \includegraphics[width=0.95\textwidth]{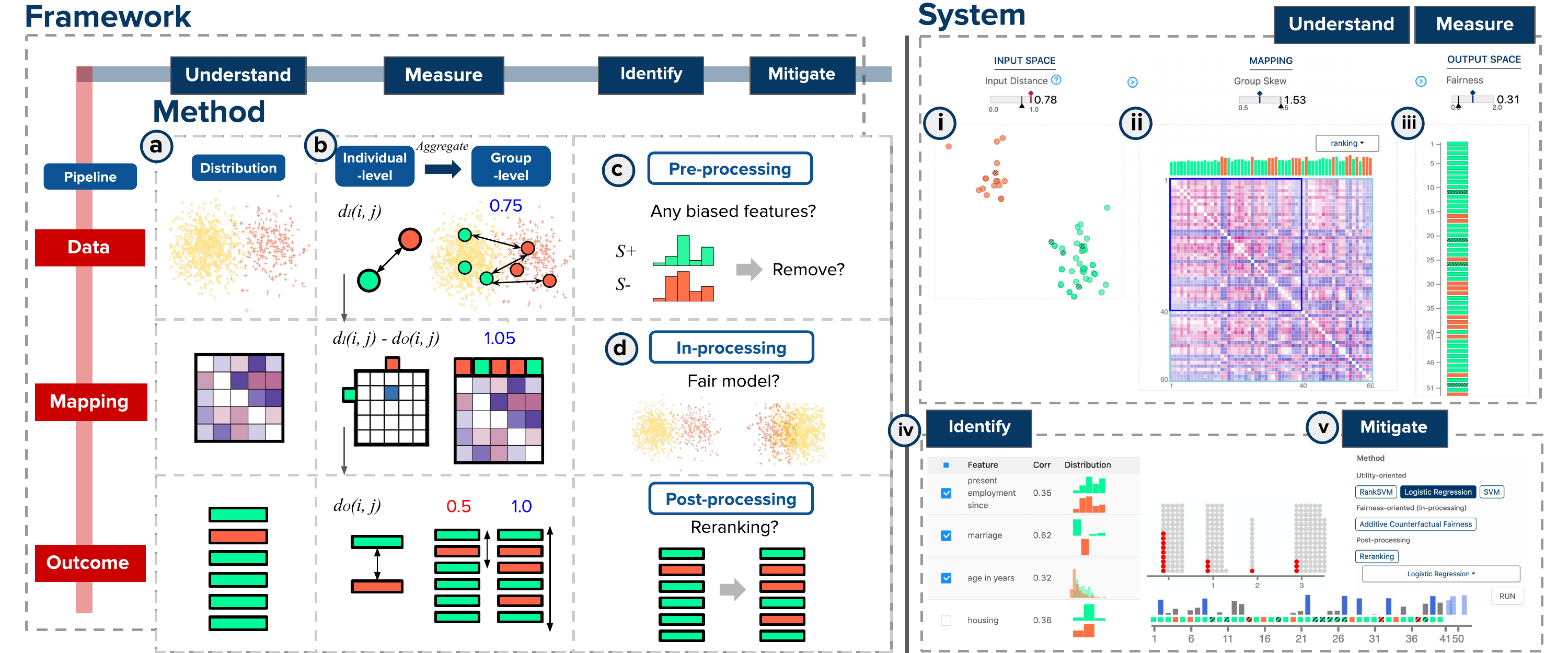}
  \vspace{-0.8em}
  \caption{We propose a design framework to protect individuals and groups from discrimination in algorithm-assisted decision making. A visual analytic system, \name, is implemented based on our proposed framework, to help data scientists and practitioners make fair decisions. The decision is made through ranking individuals who are either members of a protected group (orange bars) or a non-protected group (green bars).
  (a) The system provides a pipeline to help users understand the possible bias in a machine learning task as a {\it mapping} from the {\it input} space to the {\it output} space. (b) Different notions of fairness -- {\it individual} fairness and {\it group} fairness -- are measured and summarized numerically and visually. For example, the individual fairness is quantified by how pairwise distances between individuals are preserved through the mapping. The group fairness is quantified by the extent to which it leads to fair outcome distribution across groups, with (i) a 2D plot, (ii) a color-coded matrix , and (iii) a ranked-list plot capturing the pattern of potential biases. The system provides diagnostic modules to help (iv) identify and (v) mitigate biases through (c) investigating features before running a model, and (d) leveraging fairness-aware algorithms during and after the training step.
}
  \label{fig:teaser}
}
\vgtcinsertpkg

\ieeedoi{10.1109/TVCG.2019.2934262}
\begin{document}
\firstsection{Introduction}
\maketitle

Data-driven decision making about individuals has become ubiquitous nowadays. With the pervasive use of big data techniques, companies and governments increasingly rely on algorithms to assist in selecting individuals who meet certain criteria. In many cases, this process is conducted by first ``ranking'' the individuals based on their qualifications and then picking the top $k$ candidates based on the available resources or budgets. These ranking-based decision processes that concern rank-ordering individuals by their likelihood of success or failure have been widely adopted in many domains ranging from policing, recidivism, to job recruiting, and credit rating, which has a great impact on individuals' lives \cite{citron2014scored}. 

A critical issue of data-driven decision making is the possibility of intentionally or unintentionally discriminating against certain groups or individuals. While decision makers try to best utilize available information, including personal profiles such as race, sex, and age, the increasing cases of discrimination in the use of such personal profiles has been reported in real-world decision making. For example, a recent news \cite{reuters2018} reported that Amazon's recruiting tool, trained from a 10-year historically male-dominant resume dataset, has been found biased in favor of men. Propublica \cite{propublica2016} also reported that the recidivism scores learned from the algorithms tended to assign a higher score to an African-American defendant than to a White defendant who has been convicted of the same degree of crime. As shown in these real-world incidents, data-driven decisions are not free from existing bias. It has been pointed out that algorithmic decision making from data with inherent biases are just nothing but a systematic way of disseminating such biases to large number of people at once \cite{o2017weapons}.

Data-driven decisions are criticized not only for being biased but also for lack of explanations. Even not limited to fairness problem, recent studies are increasingly aware that a black-boxed machine learning model lacks the explanation on predicted outcome \cite{gunning2017explainable}. The machine learning models in societal decision making have assisted in judging whether individuals are qualified or not, where any results with greater performance metrics tend to be accepted without carefully examining {\it why}.

Fair and transparent machine learning in the real-world practice of decision making is in an urgent need; however, there is a lack of viable tools available to assist data science practitioners and decision makers in tackling the fairness problem. A variety of disciplines have made progress in developing fair algorithms and measures, but those are developed separately from decision-making contexts and not available in practice. While new tools became available recently \cite{googlewhatif, ibmfairness360}, none of these provide a comprehensive view and workflow to better cope with various fairness issues in the decision-making pipeline. With advancing research on measures, algorithms, and diverse perspectives on fairness, we now move one step further: to propose a viable decision-making tool to assist in fair decision making throughout the machine learning pipeline.

We argue that it is time to bring the research into real-world practice to create an impact on societal decision-making. In this paper, we present a fair decision-making framework, called \fairdm, that identifies a set of guidelines in the algorithmic-aided decision making workflow. In this work, we focus on the problem of various high-stakes decision-making process, such as credit rating and recidivism risk prediction, which involve rank-ordering individuals. Moreover, a variety of prediction problems, such as binary or multiclass classification/prediction, can be cast as a ranking problem. The proposed \fairdm is a model-agnostic framework that does not depend on a particular (ranking) algorithm, and it aims to provide a fairness pipeline to guide the examination of fairness at each step (from input to output) in the workflow. We develop \name, a visual analytic system that integrates the \fairdm framework and analytic methods into a viable tool for fair decision making. Our main contributions include: 

\begin{itemize}
    \setlength{\itemsep}{0.2pt}
    \item \textbf{Fair decision making framework (Fig. \ref{fig:teaser}).}
    We propose \fairdm framework that facilitates the contemplative decision-making process \cite{padilla2018decision}  with a set of tasks to achieve fairer decision making. Our framework incorporates the different notions of fairness (including group and individual fairness) to support understanding, measuring, identifying and mitigating bias against certain individuals and groups.
    \item \textbf{Fairness measures and methods for explainability.} We introduce a set of measures and methods to summarize the degree of bias, evaluate the impact of features leading to bias, and mitigate possible sources of bias. Our approach supports both global- and instance-level explanation for the reasoning behind the fairness of ranking decision. 
    \item \textbf{Fair decision making tool (Fig. \ref{fig:teaser}).} We develop a viable fair decision making system, \name, to assist decision makers in achieving fair decision making through the machine learning workflow. We introduce a novel representation that visualizes the phases in a machine learning workflow as different spaces where individuals and groups are mapped from one to another.
    \item \textbf{Evaluation.} We present a case study to showcase the effectiveness of our system in real-world decision-making practice. Moreover, we conduct a user study  to demonstrates the usefulness and understandability of our system in both an objective and subjective manner. Our study suggests that \name has a superior advantage over an existing tool \cite{googlewhatif}.
\end{itemize}
\section{Related Work}

\subsection{Fair Ranking}

Today's decision making has increasingly relied on machine learning algorithms such as classification and ranking methods. We mainly discuss ranking in decision making due to its broad applications. In fair ranking, early studies mainly focused on quantifying the discrimination with proposed ranking measures in the \topk list \cite{pedreschi2012study}, or indirect discrimination \cite{zliobaite2015survey}. Recently, fair ranking methods have been proposed \cite{asudehy2017designing, zehlike2017fa, karako2018using}. Asudeh et al. \cite{asudehy2017designing} scored items based on a set of desired attribute weights to achieve fairness. On the other hand, Karako and Manggala \cite{karako2018using} presented a fairness-aware Maximal Marginal Relevance method to re-rank the representation of demographic groups based on their dissimilarity as a post-hoc approach. Zehlike et al. \cite{zehlike2017fa} also proposed a re-ranking method of picking candidates from the pools of multiple groups with the desired probability. Online ranking systems, such as search engines or recommender systems, use ranking algorithms to generate an ordered list of items such as documents, goods, or individuals. The fairness problem here is to pursue the degree of exposure and attention fairly for groups \cite{singh2018fairness, yang2017measuring} or individuals \cite{biega2018equity}.

In this work, we go beyond the issue of fair exposure/attention in ranking systems and broadly consider more broadly how a system can and should best help decision makers to rank items fairly when considering the trade-offs among different notions of fairness and utility.

\subsection{Explainable Machine Learning}
Machine learning and AI approaches have been recently criticized for the lack of capability in reasoning and diagnosing the logic behind the produced decisions \cite{gunning2017explainable}. With the increasing awareness associated with this problem, explainable machine learning techniques have been proposed. A number of studies have focused on interpreting the interaction between inputs and predictions from the original model, by training a secondary interpretable model to capture instance-level \cite{ribeiro2016should} or global-level pattern \cite{angelino2017learning, frosst2017distilling}. For example, feature-level auditing methods seek to analyze the feature importance by permutation \cite{fisher2018model} or quantifying feature interaction \cite{brooks2015featureinsight, friedman2008predictive, krause2016interacting}. 
Instance-level explanations that identify instances such as counterfactual examples or prototypes \cite{kim2016examples} seek to generate an explanation with a single instance. Recent research in visual analytics integrated machine learning tools with intuitive and manipulable representation and interface. Examples include RuleMatrix \cite{ming2019rulematrix}'s rule-based visual interface for explaining decision rules based on the secondary decision tree model, the distribution-based visual representation for global-level explanation, and Rivelo's instance- and feature-level representation \cite{tamagnini2017interpreting}. Mainfold \cite{zhang2018manifold} suggested a model-agnostic framework to interpret the outcome, inspect a subset of instances, and refine the feature set or model, to facilitate the comparison of models.

None of the aforementioned approaches have addressed the explainability with respect to fairness. In this work, we leverage state-of-the-art techniques, including feature auditing, to capture the feature-induced bias at both global and instance levels. We further propose new metrics via neighborhood comparison to capture both the global- and instance-level fairness with evidence of potential unfair outcomes.

\subsection{Frameworks for Promoting Fair Machine Learning}
Fair decision-making aid is an emerging topic. With the increasing awareness of the importance of fair machine learning, a number of new tools, including API \cite{bantilan2018themis} and interface \cite{yang2018nutritional}, as well as integrated systems such as the What-if tool \cite{googlewhatif} or AI Fairness 360 \cite{ibmfairness360}, have been developed. While these tools offered a combination of explainable machine learning techniques and fair algorithms, none of them provides a comprehensive guideline to help users take proper actions to address various fairness issues throughout the machine learning decision pipeline. In contrast, \name is developed based on the \fairdm framework, with a goal to empower users with a better understanding of various potential biases and a suite of tools to identify and mitigate the biases. In our evaluation study, we compare \name with the What-if tool and demonstrate several strengths of our design.
\section{Fair Decision Making Framework} \label{sec:framework}
In this section, we present \fairdm, a decision-making framework that aims to support a better understanding of fair decision-making processes. We consider such a process as a series of required actions that decision makers need to take in order to ensure the fairness of the decision-making process and outcome is in check as shown in Fig.~\ref{fig:teaser}. We start by formulating the \topk ranking problem, followed by elaborating on the stages and the rationale behind them.

\subsection{Top-$k$ Ranking Problem}
In the framework, we assume that a decision maker requires to select the best \topk individuals in the pool of $n$ candidates $C = \{1,2,3, ... n\}$. The goal is to rank the $n$ individuals by the probability of being classified as qualified (positive) through learning a predicted value $\hat{y_i}$, where $\hat{y_i}$ represents the predicted level of qualification for an individual $i$. The learning is based on a subset of $p$ features $\mat{X}'=\{\vec{x}_1, ..., \vec{x}_p\}$ selected from full feature set $\mat{X}$.
The final outcome is an ordered list of ranked individuals $R = \langle r_i \rangle_{i\in C}$.

The following concepts will be involved when discussing the fairness in ranking decisions. A {\it sensitive attribute} is a characteristic of a group of individuals (e.g., gender or race) where, within existing social or cultural settings, the group is likely to be disadvantaged/discriminated and hence needs to be protected (as often regulated by a non-discrimination policy). We refer to a {\it protected} group, denoted as $S^+$, as a group that is likely to be discriminated in the decision-making process, and we refer to the remaining as {\it non-protected} group, denoted as $S^-$. In addition, a {\it proxy variable} is a variable correlated with a sensitive attribute whose use in a decision procedure can result in indirect discrimination \cite{datta2017proxy}.

\subsection{Machine Learning Pipeline}
We consider the decision making process as a simple machine learning pipeline consisting of three phases: \data, \model, and \outcome. The primary machine learning task in this context is to select the \topk candidates based on available features.

{\bf Data.} Given a feature set $\mat{X}$, a decision maker selects a set of features $\mat{X}'\subset \mat{X}$ to represent the qualification of candidates and seeks to learn the candidates' true property (e.g., qualified or not), denoted as a target value $y$. Each individual $i$ is represented by a set of qualification information, $\mat{X}'_i=\{\vec{x}_{i1}, ...,\vec{x}_{ip}\}$, and a target value $y_i$.

{\bf Model.} A machine learning model is a function $f(\mat{X}')$ that maps the individuals' features to the ranking outcome.

{\bf Outcome.} The outcome of the machine learning task is a ranking $R=\langle r_1, r_2, ..., r_n \rangle$, where $n$ individuals are ordered by their predicted qualification.

\subsection{Fairness Pipeline} \label{fairness_pipeline}
Given the machine learning pipeline, we propose a comprehensive workflow that consists of required actions to be supported by a fair decision making tool. For all three phases in the machine learning pipeline, a fair decision making tool should support the four fairness-aware actions: (1) \understand how every step in the machine learning process could potentially lead to biased/unfair decision making, (2) \measure the existing or potential bias, (3) \identify the possible sources of bias, and (4) \mitigate the bias by taking diagnostic actions. We provide the rationale for each action in the following. 

\subsubsection{Understand}
The first action of the fairness pipeline is to clearly understand the machine learning process and its consequences to fairness in decision making. The challenge is how to facilitate such an understanding as many practitioners do not fully recognize how every step in the process could potentially lead to biased decision making \cite{holstein_improving_2018}. To address this, we propose that a fair decision-making tool should take proactive action to help decision makers understand the possible unfairness at each machine learning stage, by providing an overview with a step-by-step workflow to guide users to examine different notions of fairness.

\subsubsection{Measure}
With an overall understanding of various fairness issues, the next step is to quantify the degree of fairness and utility and evaluate how each machine learning phase impacts fairness. While many studies working on proposing the measures primarily focus on measuring the outcome bias, we argue that quantifying bias throughout all phases of the pipeline should be made available to users to detect not only {\it consequential} bias in \outcome, but also {\it procedural} bias in \data and \model phase as well.

\subsubsection{Identify}
Upon understanding and measuring bias, decision makers need to remove the potential bias. As a crucial step to achieve this, we emphasize the importance of identifying bias from features in the dataset. In data-driven decision making, feature selection is an important step that captures information based on which individuals' qualification should be evaluated. Feature inspection tools should help identify potential bias with respect to not only the sensitive attribute but also the likely proxy variables.

In our framework, we incorporate per-feature auditing modules to investigate bias being involved in all phases of the machine learning process: pre-processing, in-processing, and post-processing bias within features.

\subsubsection{Mitigate}
Informed by the aforementioned diagnostic actions, \mitigate is where decision makers take actions to remove bias within the machine learning pipeline. We consider mitigating the bias in each of the following phases:
(1) \data/pre-processing: How does one incorporate fairness-aware feature selection?
(2) \model/in-processing: How does one select machine learning model with less bias?
(3) \outcome/post-processing: How does one adjust ranking outcome to make it fairer?

\section{Method}\label{sec:method}

Based on the \fairdm framework, we propose \name to enable fair decision making in machine learning workflow. This section presents our analytical methods to support each of four required actions (\understand, \measure, \identify, and \mitigate), and the system design will be introduced in the next section. Table. \ref{tab:fairdm_task} provides a summary of the requirements, tasks, and corresponding methods.

\begin{table*}[!ht]
  \centering
  \includegraphics[width=.95\textwidth]{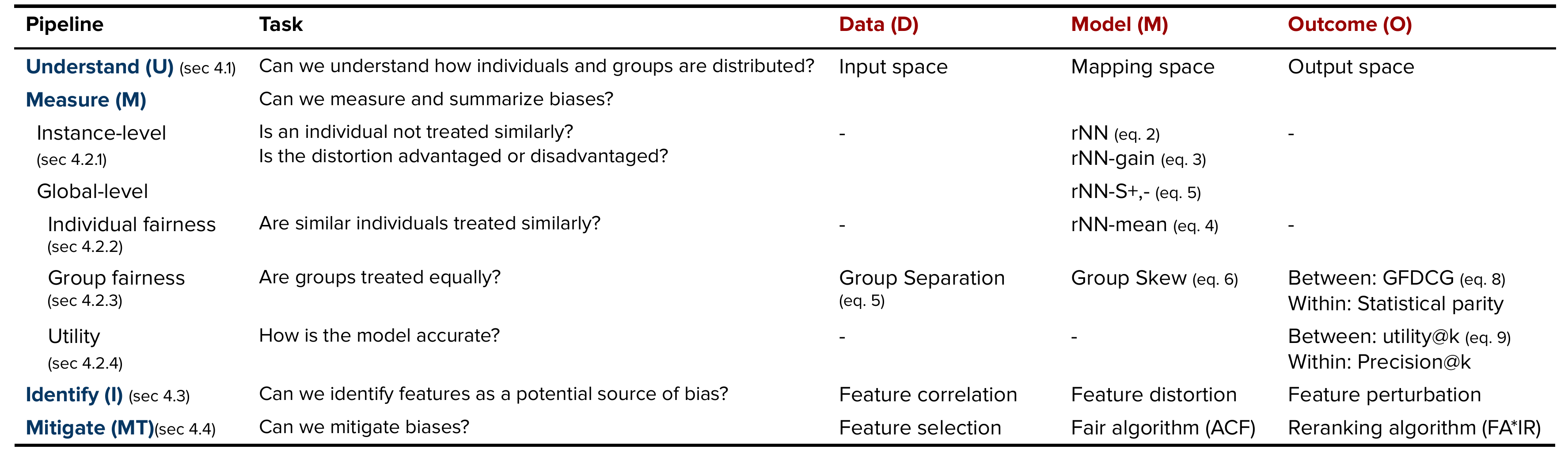}
  \caption{\fairdm tasks \& \name measures}
  \label{tab:fairdm_task}
  \vspace{-3em}
\end{table*}
\subsection{Understanding Bias}\label{sec:method_understanding_bias}

\name seeks to facilitate an effective understanding of machine learning process by introducing a novel ``\textit{space-mapping}'' representation. Inspired by Friedler et al. \cite{friedler2016possibility}, we consider the \data and \outcome phases as two {\it metric spaces} (\inputspace and \outputspace spaces), and the machine learning model as ``{\it the interactions between different spaces that make up the decision pipeline for a task}''  (Fig. \ref{fig:teaser}a). Then, biases can be introduced in each space or through the mapping between two spaces.

{\bf Input space.} We denote the \inputspace space as $\mathcal{I}=(\mat{X}, d_\mat{X})$, where $\mat{X}$ is the feature space and $d_\mat{X}$ is a distance metric defined on  $\mat{X}$. 

{\bf Output space.} The \outputspace space is noted as $\mathcal{O}=(O, d_O)$, where $O$ is an ordered list where individuals are ranked by a decision process, and $d_O$ is a distance metric defined on $O$. Both ranking and classification algorithms may be involved in such a process. When a classification model is used, the probability of each instance being classified as positive can be used to generate the ranking.

{\bf Mapping.} A machine learning model is a map $f:\mathcal{I} \rightarrow \mathcal{O}$ from the \inputspace space $\mathcal{I}$ to the \outputspace space $\mathcal{O}$.
\subsection{Measuring Bias}\label{sec:measuring_bias}

\name provides a comprehensive set of measures to support the \measure requirement (Fig. \ref{fig:teaser}b) covering the following aspects: (1) These measures are defined and organized consistently with the aforementioned the space-mapping representation (Section~\ref{sec:method_understanding_bias}). (2) It covers different (complimentary) notions of fairness, including {\it individual} and {\it group} fairness (details below). (3) The fairness individuals and groups are measured at both instance-level and global-level to enable examining the detailed and summative evidence. (4) In addition to fairness measures, we also introduce utility measures in the context of the ranking decision that allows for trade-off comparison. We first describe the metrics defined in spaces and then present the various measures.

\subsubsection{Distance and Distortion}\label{sec:distance_distortion}
Distance and distortion are two fundamental notions in \name. Given two individuals $i$ and $j$, a pairwise distance $d(i,j)$ indicates how the two individuals are dissimilar to each other in a space. When two individuals are mapped from one space to another, the pairwise distance may not be preserved. The distortion is defined as the discrepancy between two pairwise distances, indicating the degree of which the mapping is not preserved. We incorporate two different distance metrics for the \inputspace and \outputspace spaces. For \inputspace space, we calculate the Gower distance $d_\mathcal{I}(\cdot)$ \cite{gower1971general} between $\mat{X}'_i$ and $\mat{X}'_j$, which measures the pairwise distance of the two individuals' feature representations. For \outputspace space, the pairwise distance is computed using the absolute ranking difference $d_\mathcal{O}(\cdot)$ of two individuals $r_i$ and  $r_j$. Then, the distortion of two spaces is computed as the absolute difference of two pairwise distances between the \inputspace space and \outputspace space, as:
\begin{equation}
distortion_{(i,j)} = |d_\mathcal{I}(\mat{X}'_i, \mat{X}'_j) - d_\mathcal{O}(r_i, r_j)|,
\end{equation}

\subsubsection{Individual Fairness}\label{sec:individual_fairness}
By the definition of individual fairness, ``\textit{similar individuals should be treated similarly}'' \cite{dwork2012fairness}, we measure individual fairness based on the degree to which the pairwise distances in \inputspace space is preserved in \outputspace space through the mapping, i.e., based on the notion of distortion (Section~\ref{sec:distance_distortion}). \name provides both instance- and global-level examinations for individual fairness as follows. 

{\bf Instance-level bias.} Instance-level bias is measured as the amount of distortion with respect to an individual compared with other similar individuals. We capture the ``similar individuals'' of an individual $i$ based on $i$'s $h$ nearest neighbors (i.e., the closest neighbors in the \inputspace space), denoted as $NN_h$ ($h=4$ in this work).  Then, fairness with respect to an individual $i$ is measured based on how $i$ and the nearest neighbors in \inputspace space are close to each other in the \outputspace space (based on the ranking outcome), as follows:
\begin{equation} \label{eq:rnn}
rNN(i)=1-\frac{1}{h} \sum_{j \in NN_h(i)} \frac{|r_i - r_j|}{|C|}, 
\end{equation}
where the absolute ranking difference is normalized both by the number of individuals $|C|$ and the number of nearest neighbors $h$.

While $rNN$ quantifies the degree of bias/fairness with respect to an individual, the individual may be disadvantaged  (i.e., ranked much lower than the neighbors), or advantaged (i.e., ranked much higher than the neighbors). To understand the differences in bias, a signed ranking distortion for an individual $i$ is defined as: 
\begin{equation}
rNN_{gain}(i)=1-\frac{1}{h} \sum_{j \in NN_h(i)} \frac{(r_i - r_j)}{|C|}.
\end{equation}

{\bf Global-level bias.} The global-level measure of individual fairness can be obtained by aggregating (averaging) over all instance-level measures, as:
\begin{equation}
rNN_{mean}= \frac{1}{|C|} \sum_{i\in C} rNN(i).
\end{equation}

\subsubsection{Group Fairness}\label{sec:group_fairness}
Group fairness relates to equalizing outcomes across the protected $S^{+}$ and non-protected $S^{-}$ groups in \data and \outcome phase, and the mapping between two spaces. The analysis of group fairness can be richer than that of individual fairness because, by definition, individual fairness concerns ``\textit{similar treatment for similar individuals}'', which indicates the consistency in the mapping between the \inputspace and \outputspace spaces, while group fairness concerns the distribution across groups, in terms of data representation (data bias), mapping data to outcome (mapping bias), and prediction (outcome bias), as detailed below.

{\bf Data bias.} Data bias regarding group fairness seeks to uncover any bias already inherent in the input dataset. It is captured by the degree of separation between the two groups in the \inputspace space -- as ideally, the group membership should not be uncovered until revealing the sensitive attribute. Here, we adopt the symmetric Hausdorff distance \cite{hausdorff2005set}, referred to as {\it Group Separation} score, to measure the separation between the individuals from the two groups A and B:
\begin{equation}
h(A,B)=max( \tilde{h}(A,B), \tilde{h}(B,A)),
\end{equation}
where $\tilde{h}(A,B)=max_{a \in A} \{min_{b \in B}~d(a,b) \}$ is the one-sided Hausdorff distance from $A$ to $B$.

{\bf Mapping bias.} Mapping bias regarding group fairness seeks to uncover any unfair distortion between the \inputspace and \outputspace spaces at the group level. It is defined based on comparing the pairwise distortions, $distortion(i,j)$, between the two groups (for $i \in S^+$ and $j \in S^-$) against the distortion within the groups (for $i,j \in S^{(\cdot)}$). For example, when the pairwise distances between Men and Women are distorted (i.e., when greater between-group distortion is observed), the mapping has a systemic, or {\it structural} bias. We adopt the {\it Group Skew} concept \cite{friedler2016possibility} to measure such bias as:
\begin{equation}
Group~Skew = \frac{Distortion_{BTN}}{Distortion_{WTN}}, 
\end{equation}
where $Distortion_{BTN}=\sum_{i \in S^+, j \in S^-,i \ne j}distortion(i,j)$ and $Distortion_{WTN}=\sum_{i,j \in S^{(\cdot)},i \ne j}distortion(i,j)$.

{\bf Mapping bias per group.} 
The group-specific mapping bias can be quantified based on how individuals in the group receive the mapping bias (ref. Equation~\ref{eq:rnn}). It is thus defined by averaging the mapping bias of individuals in either the protected group $S^+$ or non-protected group $S^-$:
\begin{equation}
\begin{aligned}
& rNN_{S^+}= \frac{1}{|S^+|} \sum_{i \in S^+} rNN(i), \\
& rNN_{S^-}= \frac{1}{|S^-|} \sum_{i \in S^-} rNN(i).
\end{aligned}
\end{equation}

{\bf Outcome bias.} Outcome bias regarding group fairness should capture how decision outcomes are fairly distributed across groups. In the context of ranking decisions, fairness should be evaluated between different rankings (different ordered lists). Furthermore, fairness should also be evaluated based on the choice about the \topk threshold within a given ranking list (i.e., to choose the most qualified $k$ individuals from the list). A popular method for comparing rankings is $nDCG$ \cite{busa2012apple}, which involves logarithmic discount to favor items at the top ranking positions. While this method has been widely adopted in online ranking systems to reflect users' scarce attention toward only a limited few top items, in \name, our primary concern is about whether an individual being ranked on the \topk is fair or not, rather than how much attention the individual received from the ranking position. Therefore, we consider a linear rather than a logarithmic discount in order to differentiate rankings with different orders without heavily favoring a very few top items. Our measure, called {\it GFDCG} (Group-Fairness DCG) is defined based on comparing the quality of the \topk ordering in one group against another, as:


\begin{equation}
GFDCG=\frac{linear~DCG@k(S^+)}{linear~DCG@k(S^-)},
\label{eq:gfdcg}
\end{equation}
where $linear~DCG@k(S^{(\cdot)}) = \sum_{i \in S^{(\cdot)} \cap R^{(k)}} y_i \times \frac{n-r_i}{n}$,
 $R^{(k)}$ is the \topk list truncated from the whole ranking list $R$, $r_i$ is the ranking position of an individual $i$, $y_i$ is the true qualification of $i$, and $n$ is the total number of individuals in the dataset. 
 
While $GFDCG$ is useful for comparing a different threshold $k$ within the same ranking (referred as {\it within-ranking} comparison), it is less effective in comparing different rankings when $k$ is large due to the discounting effect. Therefore, to compare the fairness between different rankings, we adopt the statistical parity, which calculates the ratio of two groups without ranking discount.

\subsubsection{Utility}
In this work, we consider the utility of a ranking as the quality of the ranking. Similar to the GFDCG, the quality is evaluated based on the extent to which the \topk ordering captures truly qualified individuals, and thus a linear discount is also adopted. The utility is defined by comparing the  {\it linear~DCG} from the predicted \topk ordering against the {\it ideal} \topk ordering ({\it IDCG}):

\begin{equation}
utility@k = \frac{linear~DCG@k}{linear~IDCG@k}, 
\label{eq:utilityk}
\end{equation}
where $linear~DCG@k = \frac{}{}\sum_{i \in R^{(k)}} y_i \times \frac{n-r_i}{n}$,~$linear~IDCG@k = \frac{}{}\sum_{i \in R_{I}^{(k)}} y_i \times \frac{n-r_i}{n}$, $R^{(k)}$ is the \topk list from the predicted ranking and $R_{I}^{(k)}$ is the \topk list from the true ranking (based on individuals' true qualifications). In the same manner, we define the within-ranking utility measure by adopting the widely used information retrieval measure $Precision@k$.

\subsection{Identifying Bias}\label{sec:identify_bias}
\name provides three different strategies in each of the machine learning phases to detect possible biases due to feature selection: (1) \data: feature correlation between sensitive attribute and other features, (2) \model: per-feature impact on outliers that received most distortions from the model, and (3) \outcome: feature importance by ranking perturbation.

{\bf Feature correlation.} At the \data phase, feature correlation analysis offers a way to detect bias regarding group fairness, by removing the information that can reveal the group membership of individuals. In \name, a highly correlated feature to a sensitive attribute is detected by comparing the distributions of the feature values by groups.  If two distributions from the protected and non-protected groups are very distinct, the features are subject to be used as a proxy for the sensitive attribute. In this work, we compute the difference between the two distributions using Wasserstein distance \cite{ruschendorf2001wasserstein}, which measures the transportation cost from one distribution to another. The greater the distance, the more likely the feature can be used to distinguish the two groups and can lead to indirect discrimination.

{\bf Feature impact on outlier distortions.} At the \model phase, outlier distortion analysis measures how a feature is correlated with the overall distortion from the mapping. To compute this, a distortion distribution is first generated from the distortions of all instances. Outliers are those having greater distortions than other instances (the right tail of the distribution). For a given feature, the distance between the distortions received by the outliers and by the rest of individuals are computed to reveal the impact of the feature on the outlier distortions. Here, the Wasserstein distance is used to compute the distance between the two distortion distributions.

{\bf Feature perturbation.} At the \outcome phase, the feature importance of each feature to the fairness and utility of ranking outcome is analyzed. We adopt the feature perturbation method \cite{breiman2001random}, a widely-used feature auditing technique to evaluate the feature impact on a model. To compute this, we permute the values of a feature $\vec{x}_q$ into $\tilde{\vec{x}}_q$ from the selected feature set $\mat{X}'$ and re-train the model with $\mat{X}_{\tilde{q}}'=\big[\vec{x}_1, \ldots, \tilde{\vec{x}}_q, \ldots \big]$. The permutation is performed by swapping the feature values $\{\vec{x}_{q1}, \vec{x}_{q2}, ..., \vec{x}_{q(2/n)}\}$ with $\{\vec{x}_{q(2/n+1)}, ..., \vec{x}_{qn}\}$ as suggested by \cite{fisher2018model}. Then, the impact of the feature is measured by how much the fairness and utility measures (ref. Equation \ref{eq:gfdcg} and \ref{eq:utilityk}) drop compared with the model with the original non-permuted features.
\begin{figure*}[!ht]
  \centering
  \includegraphics[width=.93\textwidth]{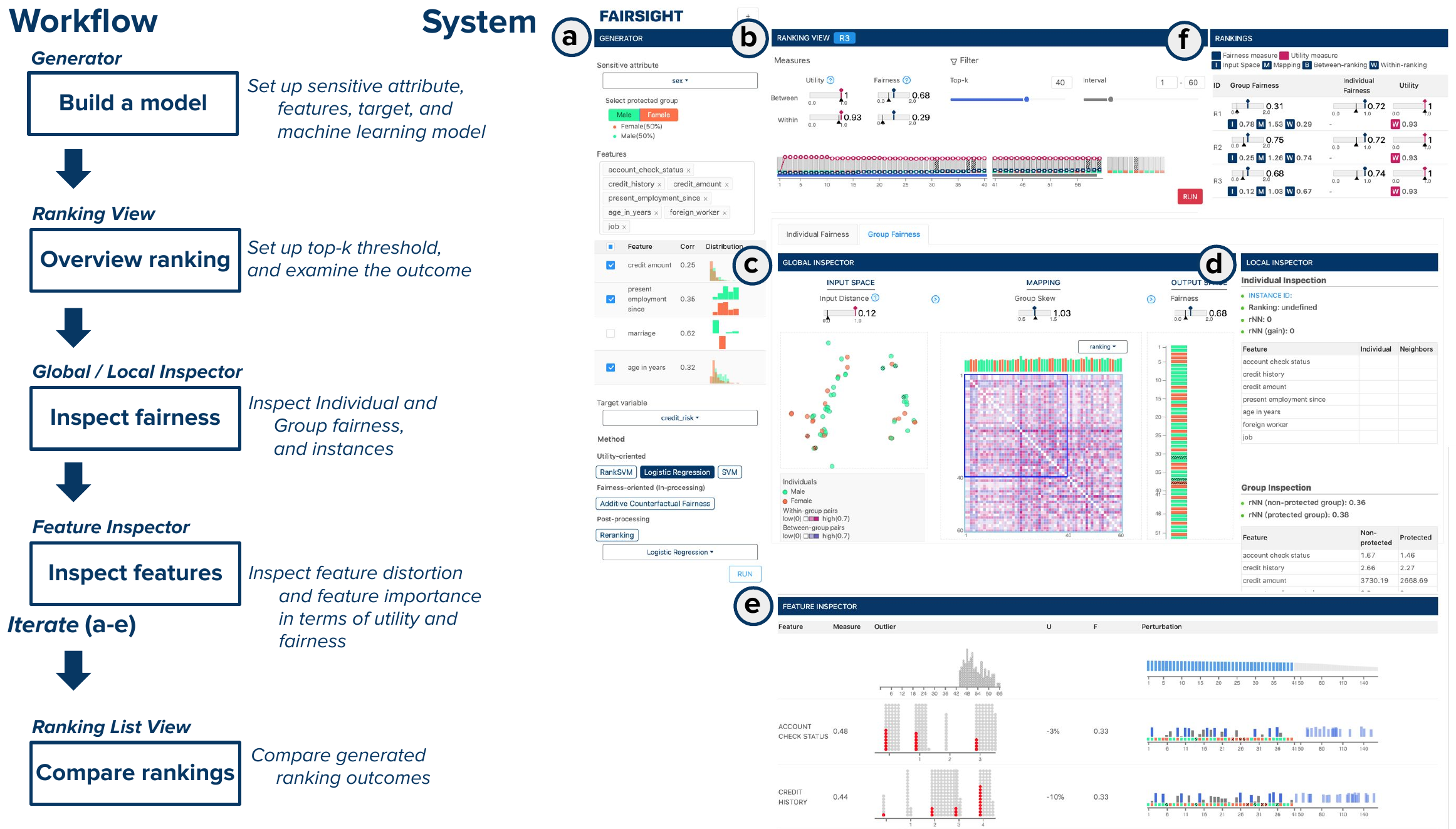}
  \vspace{-0.8em}
  \caption{The workflow of fair decision making in \name. (a) It starts with setting up inputs including the sensitive attribute and protected group. (b) After running a model, the ranking outcome and measures are represented in \rankingview. (c) \globalinspection visualizes the two spaces and the mapping process of Individual and Group fairness provided in the separate tap. (d) When an individual is hovered, \localinspection provides the instance- and group-level exploration. (e) In \featureinspection, users can investigate the feature distortion and feature perturbation to identify features as the possible source of bias. (f) All generated ranking outcomes are listed and plotted in the \rankinglist.}
  \label{fig:workflow_system}
  \vspace{-1em}
\end{figure*}

\subsection{Mitigating Bias} \label{mitigating_bias}

\name provides three types of methods for reducing biases in each of the three phases in the machine learning pipeline: pre-processing, in-processing, and post-processing methods.

{\bf Pre-processing.} Pre-processing method can be considered as a preemptive action to make the feature selection step as free from bias as possible. To achieve group fairness, decision makers should avoid using the sensitive attribute as part of the selected features. In addition, any other features that are highly correlated with any of the sensitive attributes, if used intentionally or unintentionally, can lead to indirect discrimination and should be avoided as much as possible. Such features can be detected during the \identify stage as described in Section~\ref{sec:identify_bias}.

{\bf In-processing.} This method seeks to mitigate bias by selecting fair machine learning algorithms. \name incorporates Additive Counterfactually Fair (ACF) model \cite{kusner2017counterfactual}. ACF assumes that the counterfactual probabilities of two individuals choosing from either group should be the same with respect to a given (non-sensitive) feature. We also provide a plug-in architecture in \name to allow a variety of fairness-aware algorithms to be added in the system, which can be utilized to compare multiple fair algorithms and to choose one that better mitigates the bias while having a high utility value.

{\bf Post-processing.} Here, we aim to achieve a fair ranking outcome independently of the \data and \model phases by adjusting the ranking. This approach is especially useful in situations where decision makers do not have full control of phases before the outcome.  With access to the ranking outcome, a post-processing method provides a safeguard against biases in the outcome. \name incorporates a fair ranking algorithm proposed by \cite{zehlike2017fa}, which re-ranks a ranking list by randomly choosing an individual from either of two group rankings with a predefined probability. Other re-ranking algorithms can also be included through the plug-in architecture in the system.
\section{Design Goals} 
\label{sec:design_goals}

We identify a set of requirements based on \fairdm that integrates the aforementioned methods with relevant tasks. \\ \\
\textbf{R1. Enable examining different notions of fairness in the data-driven decision with consistent visual components and interactive tools.}
\vspace{-0.5em}
\begin{description}
    \item \textbf{T1.} A fair decision making tool should enable users to select a sensitive attribute and a protected group to pursue different notions of fairness, including individual fairness and group fairness. The system also should provide an integrated interface and intuitive representation consistent with various fairness notions.
\end{description}
\textbf{R2. Facilitate the understanding and measuring of bias.}
\vspace{-0.5em}
\begin{description}
    \setlength{\itemsep}{0.2pt}
    \item \textbf{T2.} The system should help users to understand the distribution of individuals and groups to figure out whether or to what extent each phase of machine learning process is biased.
    \vspace{-0.25em}
    \item \textbf{T3.} The system should provide the degree of fairness and utility in a summary to help users to obtain fair decisions among the trade-offs.
    \vspace{-0.25em}
    \item \textbf{T4.} The system should enable the instance-level exploration to help understand the reason behind how individuals and groups are processed fairly/unfairly.
\end{description}
\textbf{R3. Provide diagnostic modules to identify and mitigate bias.}
\vspace{-0.5em}
\begin{description}
    \setlength{\itemsep}{0.2pt}
    \item \textbf{T5.} The system should support feature evaluation with respect to fairness in three machine learning phases. This task seeks the evidence of features selection in pre-processing the data.
    \vspace{-0.25em}
    \item \textbf{T6.} The system should allow users to mitigate the bias to obtain better rankings. This module also should provide the mitigating methods in all three phases to achieve the procedural fairness.
\end{description}
\textbf{R4. Facilitate the comparison of multiple rankings to evaluate ranking decisions.}
\vspace{-0.5em}
\begin{description}
    \item \textbf{T7.} The system should allow users to repeat the process to generate multiple rankings and evaluate the trade-off between fairness and utility across them.
\end{description}
\section{FairSight - System Overview}
In this section, we discuss the system architecture of \name and present the major six visual analytic components.

\subsection{Generator}
\generator allows users to set up all required inputs for fair decision making (Fig. \ref{fig:workflow_system}a) (\textbf{R1}). Users can start the setting with the selection of the sensitive attribute and protected group (\textbf{T1}). In the feature table, we provide the feature correlation measures (ref. Section \ref{sec:method}) to aid the feature selection. Each feature has two bar charts which indicate the by-group feature value distribution (e.g., orange bar chart for Male, and green bar chart for Female), and the correlation measure (i.e., how the two distributions are dissimilar to each other). Users can scrutinize how each feature has a potential to be used as proxy variable of the sensitive attribute (e.g., `marriage' of the feature table in Fig. \ref{fig:workflow_system}a -- where the distributions across groups are quite different and hence can be used as a proxy for the sensitive attribute) (\textbf{T5}).

\generator also provides a list of available machine learning algorithms, including popular classification and ranking models such as Logistic Regression, Support Vector Machine, and Ranking SVM. We also include the in-processing model (Additive Counterfactual Fairness (ACF)) \cite{kusner2017counterfactual}) and the post-processing method (FA*IR \cite{zehlike2017fa}) as a way of mitigating bias (\textbf{R3}). The system also has a plug-in architecture that allows users to flexibly include more machine learning algorithms.

\subsection{Ranking View}

\begin{figure}[!ht]
  \centering
  \includegraphics[width=.9\linewidth]{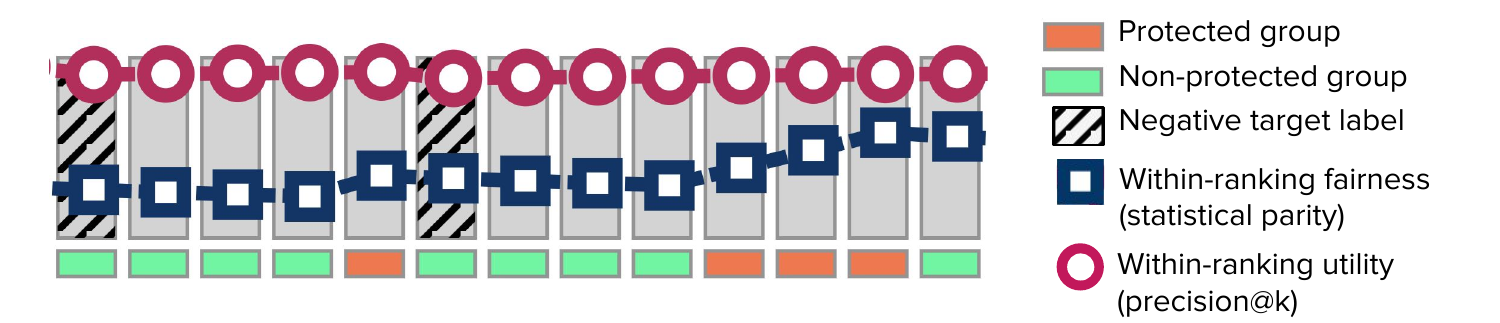}
  \vspace{-0.8em}
  \caption{The visualization of ranking outcome supports the exploration of cumulative ranking quality in terms of fairness and utility. The individuals are encoded as vertical bars with the group indicator at the bottom, with a blue rectangle (fairness) and a red circle (utility). It facilitates the selection of different $k$ values, by helping users recognize the score change as $k$ increases, as the trade-off between the fairness and utility.}
  \label{fig:ranking_view}
  \vspace{-1em}
\end{figure}

\rankingview (Fig. \ref{fig:workflow_system}b) provides an overview of the current ranking outcome. First, we report all outcome measures of fairness and utility, which includes between-ranking and within-ranking measures. While the between-ranking measures help determine whether the current ranking is better than other generated rankings, the within-ranking measures are useful to find the best \topk threshold. Along with the \topk slider is the interval slider to determine how many individuals are to be represented in \globalinspection and Feature Inspection View. \rankingview also visually represents the current ranking outcome as shown in Fig.~\ref{fig:ranking_view}. Each individual is encoded as a bar, with the group membership as a small rectangle at the bottom. A bar is filled with diagonal patch if the individual has false target label (e.g., ``not likely to pay back a loan'' in credit risk prediction)  from the target variable. On top of the bars are the trend line of within-ranking utility (dark red) and fairness (dark blue).

\subsection{Global Inspection View}
\globalinspection (Fig. \ref{fig:workflow_system}c) provides an overview of the fairness in three phases of the machine learning process to help \understand and \measure bias (\textbf{R2}). We present two notions of fairness, Individual and Group fairness, in the separate tabs to represent individuals and groups with corresponding measures independently.

This view consists of the visual components of three spaces. Each space visualizes the distribution of individuals in each phase (\textbf{T2}). The \inputspace Space View visualizes the feature representation of individuals as circles in a 2D plot using t-SNE \cite{maaten2008visualizing}. For \mappingspace space, Matrix View represents all pairs of individuals in the mapping process with the amount of pairwise distortion between two spaces.  As mentioned in Section \ref{sec:group_fairness}, two kinds of pairs (between-group and within-group pairs) are colored as purple and pink. Darker colors indicate greater distortion, as opposed to no distortion with white color). Along with the two spaces and the mapping, fairness measures of each space are presented to provide the summary of bias in each phase (\textbf{T3}).

\subsection{Local Inspection View}
\begin{figure}[!ht]
  \vspace{-1em}
  \centering
  \includegraphics[width=.85\linewidth]{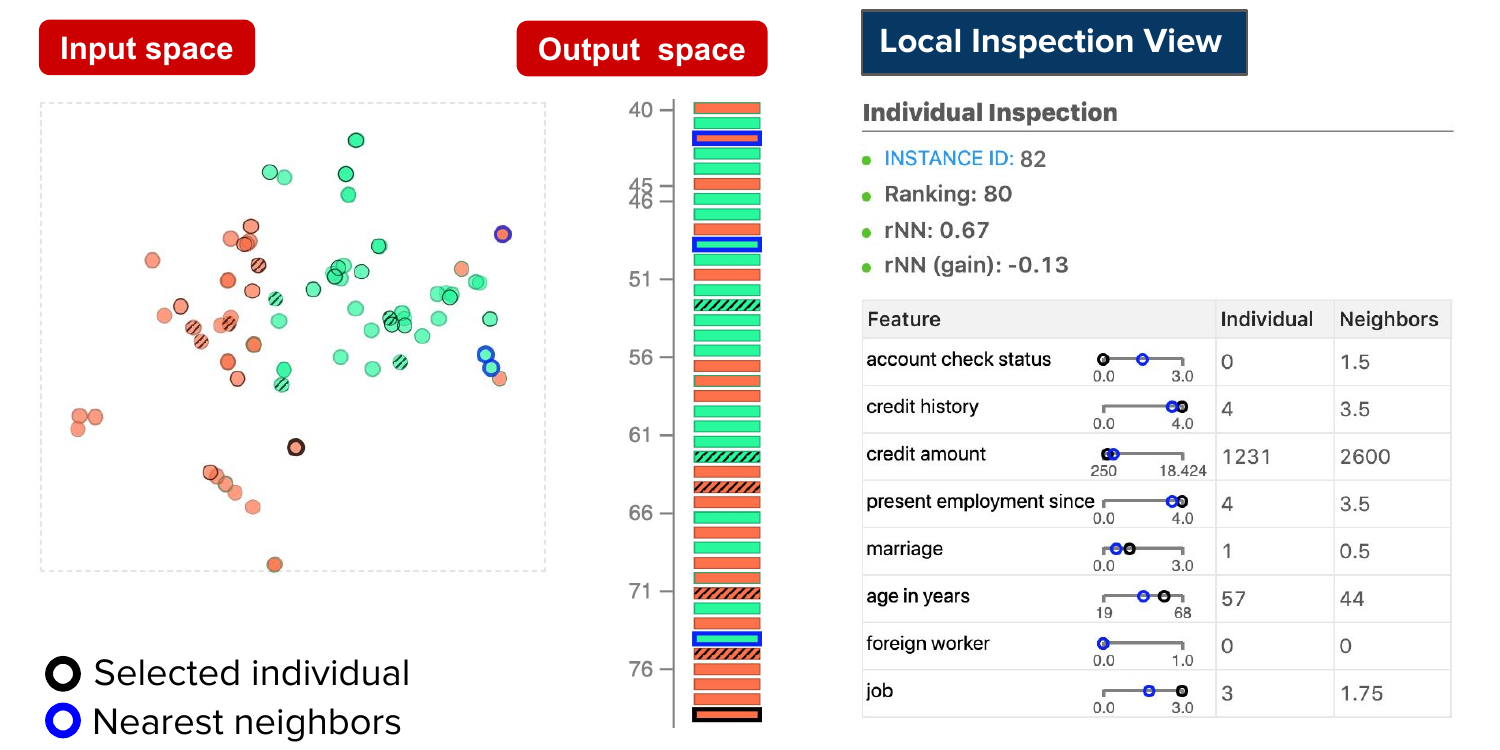}
  \vspace{-1em}
  \caption{Local inspection View. Once users hover over an instance (selected individual as black), nearest neighbors are also highlighted as blue. The feature table shows the difference of feature values (the individual's vs. the average of neighbors').}
  \label{fig:local_inspection}
  \vspace{-1em}
\end{figure}

\localinspection (Fig. \ref{fig:workflow_system}d) supports the exploration of information on a specific individual (\textbf{T4}). As shown in Fig. \ref{fig:local_inspection}, when users hover an individual in any three spaces views of \globalinspection, the individual is highlighted with black stroke, and its nearest neighbors with blue stroke. \localinspection displays the detailed information: Instance-level bias ($rNN$) and gain ($rNN_{gain}$), and feature value of the individual and its nearest neighbors. The feature table enables comparing the feature value of the individual, and the average feature value of nearest neighbors, so that users can do reasoning about what feature contributed to bias and gains. We also support the comparison of two groups. We provide group-level bias ($rNN_{S^+}$ and $rNN_{S^-}$) for users to compare which group has more bias, but also the average of feature values for each group to show the difference (Fig. \ref{fig:workflow_system}d).

\subsection{Feature Inspection View}

\featureinspection (Fig. \ref{fig:workflow_system}e) lists all selected features to support the identification of the feature bias in \model and \outcome phase (\textbf{T5}). It is composed of two components: Feature distortion and feature perturbation (ref. Section \ref{sec:identify_bias}).

For the feature distortion, we plot the overall distribution of instances with respect to their distortions. We then identify outliers that have greater distortion within 5\% of the right tail. For each feature, we represent the whole individuals (gray circle) with outliers (red circle) in a histogram along with feature correlation score. The more distinct two distributions are, with respect to certain feature, such feature is likely to be a source of bias.

\begin{figure}[!ht]
  \vspace{-0.25em}
  \centering
  \includegraphics[width=.85\linewidth]{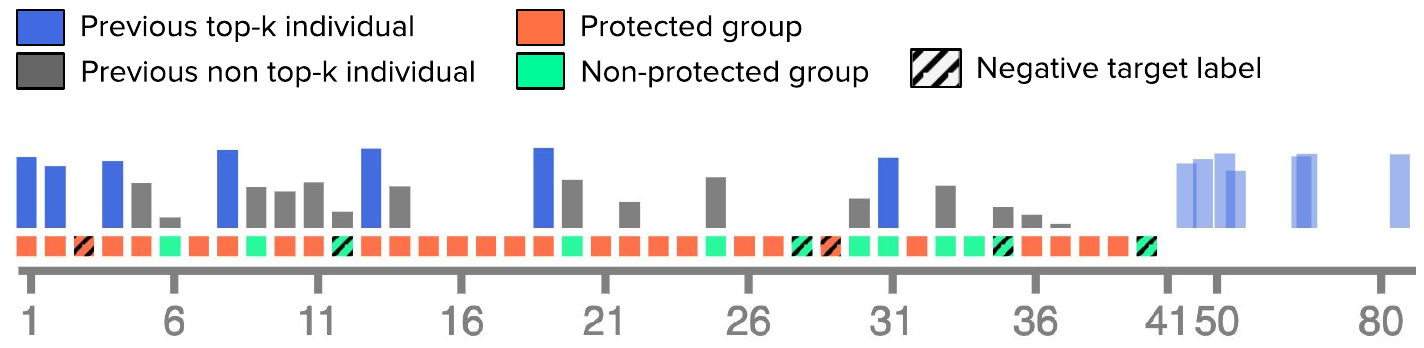}
  \vspace{-1em}
  \caption{Visual ranking component of feature perturbation. This view represents how the ranking changes after perturbation (x-axis) compared to one before perturbation (y-axis: the height of bars as previous ranking).}
  \label{fig:feature_perturbation_view}
  \vspace{-1em}
\end{figure}

Also, the result of feature perturbation for each feature is represented as the visual component of perturbed ranking in the feature table as shown in Fig. \ref{fig:feature_perturbation_view}. Each individual is represented as a bar, which is ordered by after-perturbation ranking in the $x$-axis. To represent the ranking difference by perturbation, we color the individual bars based on whether they were previously in the \topk (blue) or not (gray), and set the height as before-perturbation ranking. We also encode the information of group membership as a small rectangle (orange or green) and target label as a black striped patch (negative label). Any individuals who were in the \topk in before-perturbation ranking are represented with a semi-transparent blue bar to indicate how they are ranked.

\begin{figure*}[!ht] 
  \centering
  \includegraphics[width=.9\textwidth]{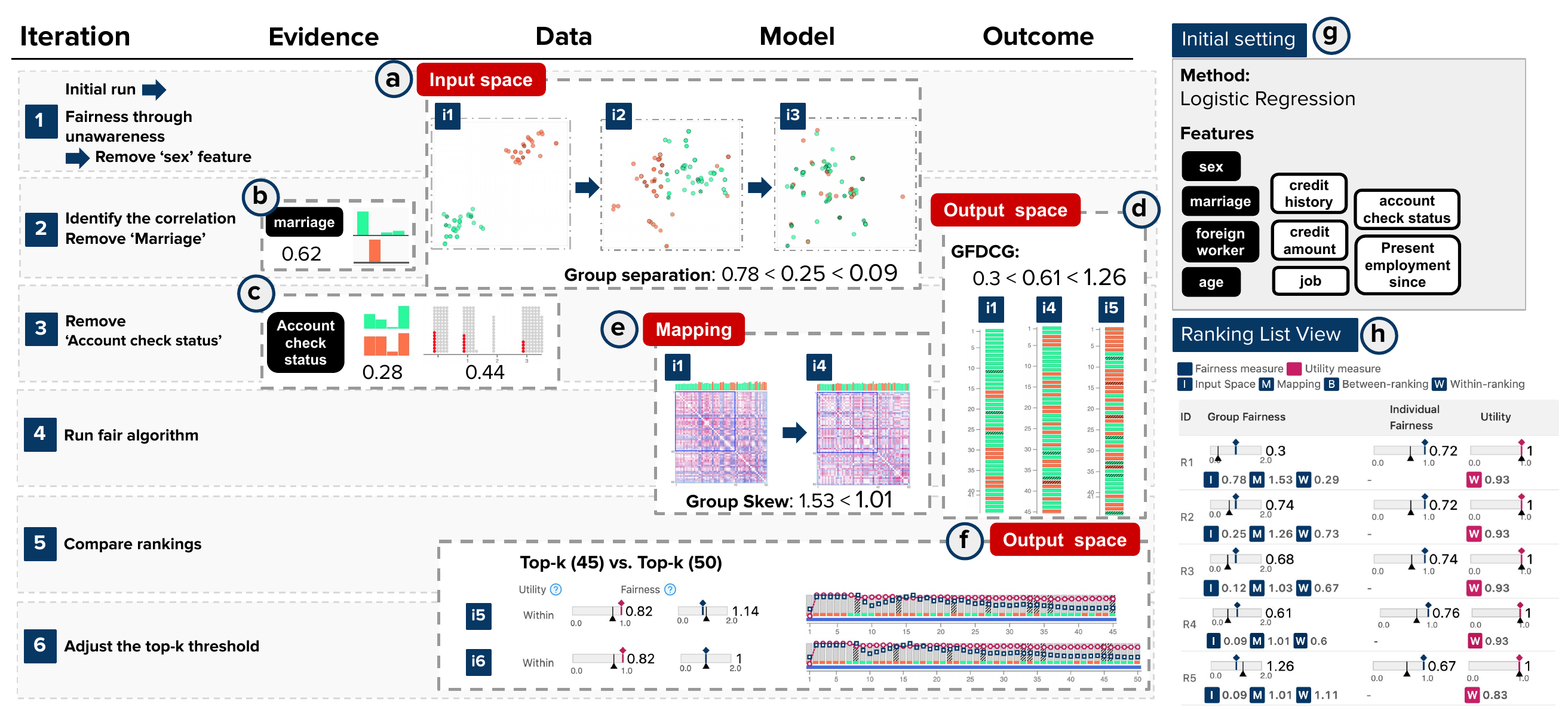}
  \vspace{-0.5em}
  \caption{Summary of case study in loan approval (g) with the initial features and method. Users can understand (a) the distribution of groups and individuals in Data phase. With the feature auditing modules to detect (b) feature correlation and (c) feature distortion, the visual components of (d) Matrix View in Mapping and (d) Ranking View in Outcome phase represent that the fairness in each phase improves. (f) Within-ranking module helps adjusting better \topk threshold. (h) \rankinglist displays the trade-off between fairness and utility to help users comparing the rankings.}
  \label{fig:case_study}
  \vspace{-2em}
\end{figure*}

\subsection{Ranking List View}
The Ranking List View provides the summary of all generated ranking outcomes as shown in Fig. \ref{fig:workflow_system}f. This view serves to compare the fairness and utility measures of rankings so that users can consider the trade-off between fairness and utility in their decision making. In the table, we list rankings which consist of Group fairness, Individual fairness, and Utility measures as columns. In particular, Fig. \ref{fig:workflow_system}f presents a number of representative measures, one for each type of fairness and utility, via multiple linear gauge plots. Each of the plots consists of an ideal score for the corresponding measure (encoded by a diamond shape) and the current score (as a triangle marker). Additional fairness or utility measures are numerically presented (Fig.~\ref{fig:workflow_system}f).
\section{Case Study}
\label{sec:case_study}
We present a case study of the loan approval decision to showcase the effectiveness of \name in achieving the fair data-driven decision (Fig. \ref{fig:case_study}). In this scenario, a credit analyst, Julia, working for a credit rating company aims to pick the best qualified $k$ customers to grant a loan. She has a fairness goal to protect female customers as the persistent discrimination against women in financing has been reported.

{\bf Settings.} We utilize the German Credit Dataset published in UCI dataset \cite{HofmannGermanCredit}. For this case study, we randomly select 250 instances with 10 features. We sample the same number of individuals from two groups (Men:Women = 5:5) and keep the ratio of target label (Credit risk: Yes or No) of each group the same as the original dataset. For the initial run, we select 9 features (Fig. \ref{fig:case_study}g) with $k=45$ as the \topk threshold. We illustrate this case by showing how Julia iteratively went over the machine learning pipeline for six iterations. In the following, we use the abbreviated notation $S_F\times S_M$ for $S_F\in \{$\understandabbr, \measureabbr, \identifyabbr, \mitigateabbr$\}$ and $S_M\in \{$\dataabbr, \modelabbr, \outcomeabbr$\}$ to highlight how each action correspond to the machine learning and \fairdm stages, e.g., \understandabbr$\times$\dataabbr~denotes the action of using tools at the \data phase to meet the \understand requirement.

{\bf Iteration 1.} Julia started the initial run with 9 features (Fig. \ref{fig:case_study}g) including ``Sex’’ feature selected with Logistic Regression model. After the model running, she realized that the ranking outcome was significantly discriminated against women ($GFDCG$ = 0.3) in Ranking View (Fig. \ref{fig:case_study}d-i1). When she took a closer look at the distribution of individuals, the ranking outcome was severely favorable towards men, especially within the top-15. In \globalinspection, all \data and \model phases were biased. Specifically, two groups were clearly separated due to inherent group bias in \inputspace space (\understandabbr$\times$\dataabbr) (Fig. \ref{fig:case_study}a-i1). In \mappingspace space, Group skew was over 1 indicating there is a structural bias (Fig. \ref{fig:case_study}e-i1). There was also a gap between the amount of bias per group as well ($rNN_{S^+}$: 0.4, $rNN_{S^-}$: 0.32). By excluding the sensitive attribute from the decision making, she deleted the ``Sex’’ feature and generated the second ranking (\identifyabbr$\times$\dataabbr).

{\bf Iteration 2.} Without ``Sex’’ feature involved, she checked that two groups are more scattered throughout \inputspace space though she could detect two groups that formed the clusters ($Group~ separation$ = 0.27) (\understandabbr$\times$\dataabbr) (Fig. \ref{fig:case_study}a-i2). There was still inherent bias, so she decided to continue examining the other potential features that brings in bias in \data phase. While investigating Feature Correlation table (\identifyabbr$\times$\dataabbr) (Fig. \ref{fig:case_study}b), she found that ``Marriage’’ feature is highly correlated to the sensitive attribute with the score of 0.62 (\measureabbr$\times$\dataabbr). The distribution plot showed that almost all men are in single status, whereas most women are in the status of married or divorced. She judged that ``Marriage’’ feature could be a potential proxy variable to discriminate against certain group. 

She also noticed in \inputspace Space that there is an individual (Woman, ID: 82) plotted in a distance from other female individuals (Fig. \ref{fig:local_inspection}). When she hovered over the individual, she found that the individual had the distortion ($rNN = 0.67$), but it turned out to be the disadvantage against the individual ($rNN_{gain}$ = -0.13). She investigated the feature value table to see how the individual is different from neighbors. She noticed that the individual was significantly different in ``Account check status'', ``Marriage'', and ``Job''. Especially, the individual had a significantly different account check status (0: No account), and was in ``Married'' status (Marriage = 1) compared to the average of neighbors' marriage status 0.5, which is closer to ``Single'' status (0: Single). This instance-level exploration enabled her to explore how each individual was biased or disadvantaged, with the difference of feature values explained. Taking all pieces of evidence from this iteration, she decided to remove ``Marriage'' feature (\mitigateabbr$\times$\dataabbr). 

{\bf Iteration 3.} She instantly noticed that removing the ``Marriage’’ feature improved the fairness score. For \inputspace space, Group separation score dropped to 0.12 (\measureabbr$\times$\dataabbr) (R3 in Fig. \ref{fig:case_study}h). In \inputspace space, she found that two groups were not clearly separated by their cluster anymore (Fig. \ref{fig:case_study}a-i3). She also found that Group Skew score in \model phase improved by 0.05. But GFDCG score in \outcome phase was still severely biased towards men, which still left much room to improve (score = 0.53). At this time, she observed in \featureinspection that ``Account check status’’ feature had the highest feature distortion (score = 0.44) and high feature correlation bias (score = 0.28) (Fig. \ref{fig:case_study}c). She finally decided to remove this qualification feature.

{\bf Iteration 4.} She immediately found that  Group Separation and Group Skew score improved by 0.03 and 0.02 (From R3 to R4 in Fig. \ref{fig:case_study}h), and Individual fairness slightly increased by 0.02, but GFDCG score was still dragged around 0.6. After all feature exploration, she decided to finalize the feature set with 7 features and run the fair method to make an improvement (\mitigateabbr$\times$\modelabbr).

{\bf Iteration 5.} When she ran the in-processing model, it improved the fairness of ranking outcome ($GFDCG$ = 1.26) (Fig. \ref{fig:case_study}d-i5) with a fair number of women within the top-20 shown in \rankingview. Finally, she found that the overall fairness score highly improved in two spaces and also Mapping ($Group~Separation$ = 0.09, $Group~skew$ = 1.01, $GFDCG$ = 1.26). The amount of biases per group was also closer to each other ($rNN_{S^+}$: 0.34, $rNN_{S^-}$: 0.33). In \rankinglist, she was able to compare all generated rankings with Group fairness, Individual fairness, and Utiliy measures. While there was a trade-off between rankings (R5 in Fig. \ref{fig:case_study}h), the last ranking outcome achieved both higher fairness and utility scores.

{\bf Iteration 6.} While she settled down with 5th ranking, she had to decide how many individuals she should pick. As she moved on to \rankingview, the ranking visualization conveyed the information of within-ranking fairness and utility trend (Fig. \ref{fig:case_study}f). Observing the nearby positions, she found that within-ranking fairness improved by 0.14 when she slightly increased the threshold to $k=50$ while within-ranking utility remains the same as 0.82. She decided to finalize the ranking decision with the last ranking by selecting 50 candidates based on the \topk threshold.
\section{User Study}
We evaluate \name's design in terms of its understandability and usefulness in decision-making by conducting a user study. We compared \name with 'What-if' tool \cite{googlewhatif} developed by Google, which is one of existing tools for probing the machine learning models on their performance and fairness.

We recruited 20 participants (age: 23–30; gender: 8 female and 12 male participants), a majority of which were graduate students who study Information and Computer Science and have the knowledge of machine learning to ensure they are familiar with the typical machine learning workflow and terminologies. We conducted a within-subject study where each participant was asked to use both tools in a random order. We gave participants the tutorial (25 mins) and let them explore the two systems (15 mins).

{\bf Questions and tasks.} Participants were given tasks in a decision-making situation similar to our case study (Section~\ref{sec:case_study}) based on the German credit rating dataset \cite{HofmannGermanCredit}. The task is to predict which candidates are most likely to pay back a loan, with a fairness goal of protecting female customers against discrimination. Due to the differences in the two systems' output decisions (i.e., FairSight: ranking; What-if: classification), the participants were asked to conduct the task differently where (1) with FairSight: to select $k$ qualified customers among $n$ candidates ($n=250$), while (2) with What-if: to classify qualified customers. Participants were asked to start with seven out of the ten features (Fig. \ref{fig:case_study}g) with Logistic Regression as initial setting.

Participants were asked to conduct 12 sub-tasks (4 fairness stage x 3 machine learning phases). These tasks correspond to the tasks listed in Fig.2 but with more specific question that has a correct answer, e.g., \mitigateabbr×\outcomeabbr: “Can you quantify the degree of fairness in the ranking outcome?”. The users were expected to correctly identify the directly relevant information offered by the system (e.g., the answer to this question could be ``a fairness score of 0.85''). We also asked users 3 additional questions for \decision (e.g., ask users to compare the differences in two iterations) and \explain (e.g., ask users to find the explanation on instances or features). The accuracy was measured based on whether a user can correctly point out the directly relevant information. We also asked users to rate the understandability ``{\it How well does the system intuitively capture and represent the bias/fairness in the decision process}?'' and usefulness ``{\it How is the information provided by system useful for fair decision-making}?'' in a Likert scale from 1 to 5 for each task.
Since the two tools have different functionality (What-if tool is able to support 9 out of 15 tasks while FairSight provides all the functionality), we measured and compared the accuracy on 9 questions which can be answered by both systems. We also collected the subjective feedback after completing the tasks.

\begin{figure}[!ht]
  \vspace{-1em}
  \centering
  \includegraphics[width=.9\linewidth]{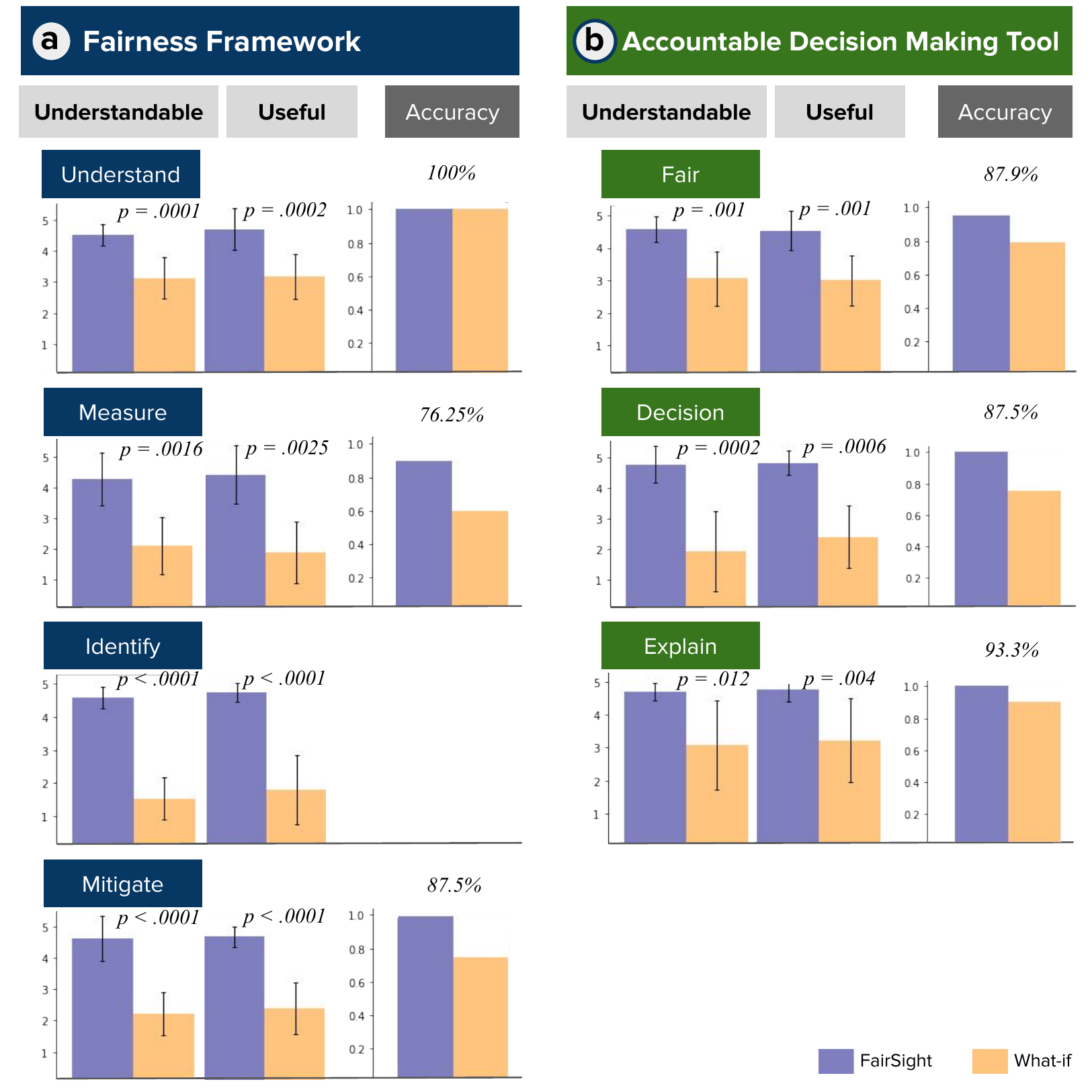}
  \vspace{-1em}
  \caption{Subjective ratings (understandability and usefulness) and accuracy  (a) in the four stages of the fairness framework and (b) in three criteria of the decision making tool: Fairness, Decision, and Explain.}
  \label{fig:user_study}
  \vspace{-1.5em}
\end{figure}

{\bf Result.} The result is presented in Fig. \ref{fig:user_study}. The overall accuracy of two tools was 95\% for FairSight, and 80\% for What-if. Per-criteria accuracy with their average accuracy is shown in Fig. \ref{fig:user_study}. Fig.~\ref{fig:user_study}a summarizes the evaluation result of each stage of fairness pipeline. The result, based on the t-test, indicated that \name significantly outperformed the What-if tool in terms of understandability and usefulness (Fig.~\ref{fig:user_study}a).

We also measured the result when ratings are aggregated by three criteria: \fair, \decision, \explain, as shown in Fig. \ref{fig:user_study}b. The statistical test proved that FairSight was more effective in terms of understandability and usefulness (Fig.~\ref{fig:user_study}b). What-if tool was relatively good at providing reasoning behind instance-level inspection using counterfactual example, and feature importance with partial dependence plot, with the score of 3.5, but lacked in comparing multiple outcomes.

{\bf Subjective feedback.} We gathered the subjective feedbacks from users. Those are summarized in three aspects: (1) \fairdm as a guideline of increasing the awareness of fairness, (2) Visualization as a fairness-aware overview of machine learning task, and (3) Comprehensive diagnosis of discriminatory features. First, most of participants appreciated how the framework and system enhanced the understanding and awareness of fairness. Several participants provided feedback on how the system improves their awareness, e.g., ``{\it It was the first time I recognized/learned how machine learning can result in the fairness problem.}'' Second, 
\name with visual components not only served as a diagnostic tool for fairness, but also helped users understand how the distribution of individuals changes with the fairness improved in three machine learning phases, as mentioned by a user, ``\textit{Three space views are intuitively show how the process is going with the degree of fairness}''.
Lastly, most of the users were surprised by how the system supports evaluating features as possible sources of direct or indirect discrimination in each phase. As a user mentioned, ``\textit{Feature selection is sometimes arbitrary, but it provides the feature-level measures as evidence of fairness-aware decision.}'' -- this demonstrated how the system can help decision makers to achieve fair decision making through better explainability.

\section{Discussion}
In this section, we discuss our observations on the use of \name, and extend it to the general context of fair decision making. We also summarize the limitations of our system based on findings from the user study.

{\bf Importance of pre-processing stage.} Although all stages of \fairdm have an important role in achieving fair decision making, the most critical part was found to be the pre-processing stage. As shown in the case study, the first 4 iterations were primarily concerned with the pre-processing stage, where the fairness scores can be significantly improved. We also observed that participants in the user study spent 80\% of their exploring sessions in detecting and removing bias from features. It is also the case of the real-world practice that, according to \cite{grgic2016case}, the data collecting and processing is the most preferable and time-consuming task. Based on our study, a fair decision tool that simply offers a package of fairness-aware algorithms and outcome measures will be not sufficient to meet the needs of data scientists and practitioners to combat the various bias and fairness issues in real-world practice, and our proposed design helps address this challenge with comprehensive support at the pre-processing stage.

\begin{figure}[!ht]
  \vspace{-0.2em}
  \centering
  \includegraphics[width=.8\linewidth]{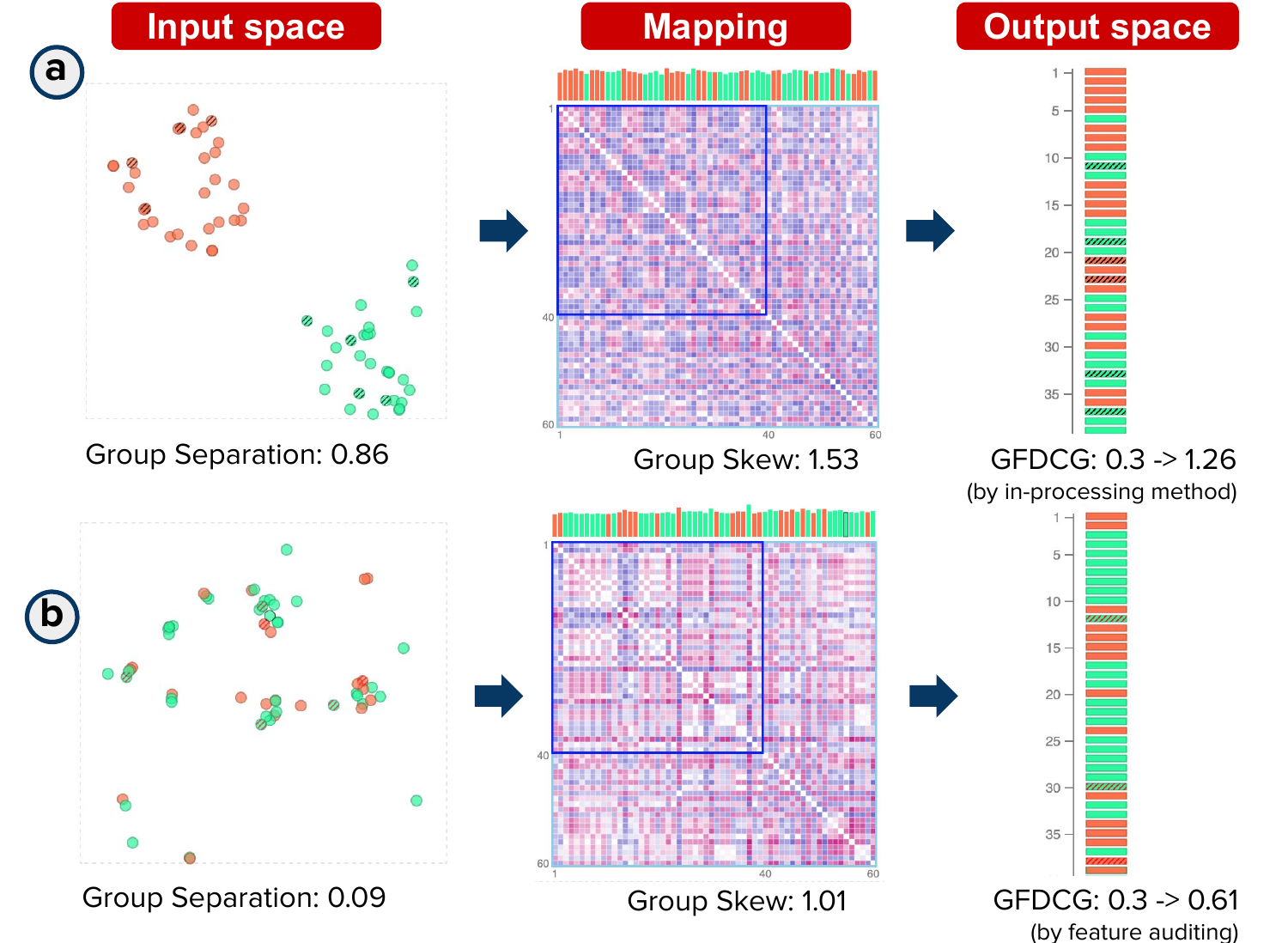}
  \vspace{-1em}
  \caption{Visualization of three stages from (a) 9 features (Iteration 1 in Case study) with the in-processing method (b) 6 features (Iteration 5 in Case study) with Logistic regression.}
  \label{fig:discussion}
  \vspace{-1em}
\end{figure}

{\bf Interaction between spaces.} \name represented the machine learning process as the mapping between \data and \outcome. With the perspective of space, our observation is that the entire pipeline coordinates in such a way bias is reinforced from data to outcome through the mapping. As illustrated in Fig. \ref{fig:discussion}, features without the pre-processing step create more bias in the mapping (Fig. \ref{fig:discussion}a), whereas features from fairer data representation were found to be less biased in the mapping and ultimately resulted in greater fairness in the ranking outcome (Fig. \ref{fig:discussion}b).

{\bf Subjectiveness of feature selection.} While it is possible to identify feature-level bias through the well-defined metrics provided by our system, human scrutiny through the interactive visualization is still required. Feature selection often requires domain knowledge; determining how a feature is important and fair may differ across contexts and domains \cite{lipton2017doctor}, and is also subjective to people's perception on fairness \cite{gower1971general}. There is no generally acceptable criteria for evaluating the trade-off between fairness and utility over decision outcomes. Therefore, it is desirable to have a decision-making tool that helps incorporate the domain knowledge and human judgment to achieve fair decision making.

{\bf Limitation.} Despite the comprehensive framework and system implementation in our study to go towards fair decision making, we observe that a few drawbacks still exist. First, our visualization creates visual clutters as a number of instances increase while it enables the instance-level exploration. Second, the visualization can be misleading depending on group population. For example, in the case when the sensitive attribute has a skewed ratio of two groups (e.g., Men:Women = 8:2), the visualization of linearly ordered ranking outcome may look unfair even for a fair ranking. Our system also treats the sensitive attribute as dichotomy between protected group and non-protected group, which may not fit into some cases.

\section{Conclusion}

In this work, we presented \fairdm, a decision making framework to aid fair data-driven decision making. We also developed \name, a visual analytic system with viable methods and visual components to facilitate a real-world practice of comprehensive fair decision making. The evaluation of our system demonstrated the usefulness of the system in fairness work over the existing tool. In the case study, we illustrated how the system was effective to measure and mitigate bias using a well-known dataset. 
For future work, we plan to extend the current binary representation of sensitive attribute in \name to handle multiple groups and sub-groups, as well as user-defined groups. Furthermore, to tackle industry-scale dataset, we will develop a scalable visual representation of rankings (e.g., how to make the matrix representation or reordering to efficiently present the fairness).

\section*{Acknowledgement}
The authors would like to acknowledge the support from NSF \#1637067 and \#1739413.

\bibliographystyle{abbrv-doi}

\bibliography{reference}
\end{document}